\documentclass[12pt]{iopart}

\usepackage[active]{srcltx} 
\usepackage{color}

\usepackage{graphicx}
\usepackage{subfigure} 
\usepackage{dcolumn}
\usepackage{bm}
\usepackage[normalem]{ulem}
\usepackage{color}

\begin{document}

\title{Zero-temperature phase diagram of D$_2$ physisorbed on graphane}  

\author{C Carbonell-Coronado $^1$, F De Soto $^1$, C Cazorla $^2$, J Boronat $^3$ and MC Gordillo $^1$}
\address{$^1$ Departamento de Sistemas F\'{\i}sicos, Qu\'{\i}micos y Naturales, Universidad Pablo de Olavide, Carretera de Utrera
, km 1, E-41013 Sevilla, Spain}
\address{$^2$ Institut de Ci\`encia de Materials de Barcelona, (ICMAB-CSIC) Campus UAB, E-08193, Bellaterra, Spain}
\address{$^3$ Departament de F\'{\i}sica i Enginyeria Nuclear, Universitat Polit\`ecnica de Catalunya, B4-B5 Campus Nord, 08034 Barcelona, 
Spain}
\ead{ccarbonellc@upo.es}

\begin{abstract}
We determined the zero-temperature phase diagram of D$_2$ physisorbed on  
graphane using the diffusion Monte Carlo method. 
The substrate used was C-graphane, an allotropic 
form of the compound that has been experimentally obtained through 
hydrogenation of graphene. We found that the ground state is the $\delta$ phase, a commensurate
structure observed experimentally when D$_2$ is adsorbed on graphite, and not
the registered $\sqrt 3 \times  \sqrt 3$ structure characteristic of H$_2$ on the same
substrate.  
  
\end{abstract}


\maketitle

\section{INTRODUCTION}
 
In recent years, we have seen an exponential growth of the 
interest in low dimensional 
forms of carbon, such as carbon nanotubes \cite{Iijima} or graphene.~\cite{science2004,pnas2005} 
Both structures are closely related to
graphite, whose upper surface has proved itself a good adsorbent for quantum gases.~\cite{colebook} 
One of the (sometimes unstated) goals of the experimental studies of quantum gases (particularly H$_2$) on relatively weak 
substrates (such as graphene versus graphite) is to find novel quasi two dimensional stable phases,
for instance, a liquid H$_2$ (or He) superfluid film all the way to T=0 K. Since this hope has not been 
fulfilled so far, new substrates have been searched to be tested.         

One of those new two dimensional substrates is called graphane, an hydrogenated version of graphene
predicted to be stable ~\cite{sofo,wash}, and one of whose forms  (C-graphane)
has been experimentally obtained
~\cite{grafanoscience}.
In C-graphane, every carbon atom is covalently bound 
to three other atoms of
the same type, and to an hydrogen atom  that sticks out perpendicularly
from the two-dimensional carbon scaffolding. Neighboring carbons  have
their bound hydrogens pointing to opposite sides of the carbon
structure. 
Hydrogen atoms  on the same side of the carbon structure are 
exactly on the same plane,
something that it is not true of all the atoms in the carbon skeleton.
Therefore, the upper solid substrate  
(the sheet of atomic hydrogen)   is
less dense than in the 
graphene case.  It has also a different symmetry:  H atoms form a
triangular lattice instead of
the hexagonal one characteristic of  graphene and graphite. However, the
underlying carbon structure, whose symmetry is still hexagonal, is close
enough to the atomic hydrogen surface to exert a sizeable influence (the
C-H length is $\sim$ 1 \AA) on any possible  adsorbate. In any case, this
novel substrate is different enough to graphite and graphene as to have been already considered as an adsorbent for
helium \cite{reatto} and H$_2$ \cite{JLTP_ccc}. In the first case, 
computer simulations predicted the ground state of $^4$He to be a liquid, 
not a commensurate solid as in the case of graphene and graphite \cite{yo}. 
On the other hand, the phase diagram of H$_2$ on C-graphane is similar to those
calculated for graphene \cite{yo2}, and found experimentally on graphite \cite{yo2,frei1,frei2,deu3175}.
In all three cases, the H$_2$ ground state is a standard $\sqrt 3 \times \sqrt 3$ solid.      

In this work, we determine the
phase diagram of D$_2$ physisorbed on top
of C-graphane. The phase diagram of D$_2$ on graphene and graphite has already 
been calculated \cite{yo3}, and found to contain different phases that those of 
H$_2$ on the same substrates. The accuracy of the results on graphite compares 
favorably against experimental results \cite{deu3175}. Then, we used similar theoretical tools 
with D$_2$ on graphane, to see if we can found significant enough differences
between the results obtained and those on H$_2$ on graphane \cite{JLTP_ccc} and D$_2$ on graphene \cite{yo3}. 
In the next Section, we 
will describe the diffusion Monte Carlo (DMC) method used 
to obtain the $T= 0$K equilibrium phases of D$_2$, giving all the
necessary information to perform the quantum calculations. The results
obtained will presented in Section III, and we will end up with the some conclusions in Section IV. 
 
\section{METHOD}

\begin{figure}
\centering
\subfigure[Potential cut at $z=3.264$ $\AA$.]{\includegraphics[angle=-90,width=8cm]{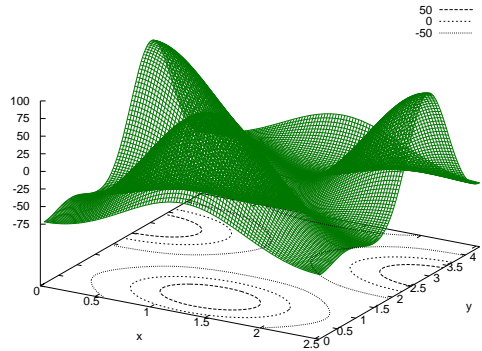}}\\
\subfigure[Wave function cut at $z=3.264$ $\AA$.]{\includegraphics[angle=-90,width=8cm]{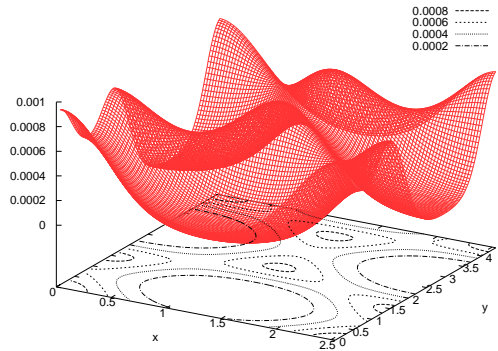}}
\caption{Numerical solution of the Schr\"odinger equation for one $D_2$ molecule in the C-graphane potential. Only the basic unit cell of
the C-graphane is 
represented.}\label{cuts}
\end{figure}

The diffusion Monte Carlo (DMC) method allows us to obtain the exact 
ground-state properties of a many-body Bose system,
such as a set of  \textit{ortho}-D$_2$
molecules adsorbed on C-graphane.  
It allows us to solve 
stochastically the $N$-body Schr\"odinger equation in imaginary time by
implementing a random walk with Gaussian and drift movements and
a weighting scheme called branching. The drift term derives from the
introduction of an importance sampling strategy through a guiding wave
function $\Psi$ (the so-called {\em trial function}), which avoids the sampling of walkers in low-probability
regions. Proceeding in this way, the variance is reduced significantly
without affecting the exactness of the results.~\cite{boro94} In practice,
the guiding function is also used to set the thermodynamic phase of the ensemble 
of particles. We will consider here a liquid phase and several solid arrangements (commensurate
or incommensurate with the substrate underneath).  
For the study of the liquid phase we used as a {\em trial function}: 
\begin{equation}
\Psi_L({\bf r}_1, {\bf r}_2, \ldots, {\bf r}_N)  =  \prod_{i<j} \exp \left[-\frac{1}{2}
\left( \frac{b}{r_{ij}} \right)^5
\right] \prod_i \Phi({\bf r}_i) \ , 
\label{trial1} 
\end{equation}
where the first term is
a Jastrow wave function that depends on the
distances $r_{ij}$  between each pair of D$_2$ molecules. The one-body term
$\Phi({\bf r}_i)$ is the result  of solving numerically the
three-dimensional Schr\"odinger equation for a molecule
interacting 
with all the individual atoms of the graphane surface. In figure \ref{cuts} we plotted a xy-plane cut of both the C-D$_2$ potential close to the potential minimum and the corresponding value for the one-body part 
of the {\em trial function}. 
During the Monte Carlo simulations, instead of
recalculating analytically both potential and wave function each time the position of
particle  {\bf r}$_i$ changes, we tabulated  $\Phi({\bf r})$
 using  a grid and then interpolated linearly
for the desired values.  Since the graphane
structure is a  quasi two-dimensional solid, it was enough to consider only the
minimum units that can be replicated in the $x$ and $y$ directions to
produce the  corresponding infinite sheet.  In our case, these units
contained eight atoms (four carbons and four hydrogens) each, and were 
chosen  to be rectangular instead of the smaller oblique cells deduced
directly from the symmetry of  the compounds.~\cite{sofo,wash}
The dimensions of this  basic unit are 2.5337 $\times$ 4.3889 \AA$^{2}$.
For the sake of comparison, the dimensions of a similar rectangular
cell for graphene are 2.4595 $\times$ 4.26 \AA$^{2}$.
The transverse displacement between neighboring carbon atoms 
in the graphane structure was
$0.46$ \AA, in agreement with \cite{wash}. 
If the position of any deuterium molecule in the
simulation cell is located outside that minimum cell,  the value of the
function $\Phi$ is obtained by projecting back that position 
within those cell limits. The grid to calculate $\Phi$  extended up to 
12 \AA~ in the  $z$ direction from the positions of the upper carbons.

The $b$ parameters of the corresponding Jastrow functions that appear in
(\ref{trial1}) were obtained from variational Monte Carlo calculations that
included ten deuterium molecules on a C-graphane simulation cell of
dimensions 35.47 $\times$ 35.11 \AA$^2$.  This is a 14$\times$8 supercell
of the basic unit defined above. The
optimal value is
$b$= 3.195 \AA, exactly the same number as the one used
for graphene in previous calculations~\cite{yo3}. Some other tests
made for different deuterium densities left the parameter unchanged.

To simulate solid deuterium phases, we multiplied  
$\Psi_L$ (\ref{trial1}) by a product of Gaussian functions whose role is to  
confine the adsorbate molecules around the
crystallographic positions ($x_I,y_I$) of the two dimensional solids
we are interested in.  We have used the Nosanow-Jastrow model,
\begin{equation}
\Psi_{S}({\bf r}_1, {\bf r}_2, \ldots, {\bf r}_N)  = \Psi_L \,  \prod_{i,I=1}^{N}
\exp\{-c[(x_i-x_I)^2+ (y_i - y_I)^2]\} \ ,
\label{trialsol}
\end{equation} 
where the $c$ parameters are dependent on the particular solid, commensurate or incommensurate. 
The variationally optimized values for $c$ are given in 
table \ref{c}.   
For the triangular incommensurate structures, the values listed are the ones for 
densities $\rho$ = 0.11 \AA$^{-2}$ and $\rho =$ 0.08 \AA$^{-2}$. A linear
interpolation was used for
intermediate adsorbate densities.

\begin{table}
\caption{\label{c}Optimal values of $c$ parameters in (\ref{trialsol}).}
\begin{indented}
\item[]
\begin{tabular}{ccc}
Phases& c (\AA$^{-2}$)\\
\hline
$\sqrt3\times\sqrt3$&0.53\\
$\delta$&0.82\\
$\epsilon$ &1.02\\
4/7&2.38\\
7/12&2.74\\
Incommensurate solid &3.1  $^ {\rm a}$\\
&1.1  $^{\rm b}$\\
\end{tabular}

\item[]$^ {\rm a}$ {For a density of 0.11\AA$^{-2}$.}
\item[]$^ {\rm b}$ {For a density of 0.08\AA$^{-2}$.}

\end{indented}
\end{table}

An important issue in the microscopic description of the system is the choice
of the empirical
potentials between the  different species involved that enter in the Hamiltonian. 
The deuterium-deuterium interaction  
was the standard of Silvera and Goldman,~\cite{silvera}, that depends only on the 
distance between the center-of-mass of each pair of hydrogen molecules. This is clearly 
an approximation, since neither the H$_2$ molecule nor the D$_2$ one have perfect spherical 
symmetry. However, the differences between the ideal spheres and the real ellipsoids
are small enough to reproduce accurately the experimental bulk phase diagram of H$_2$ at low pressures
\cite{bonin2013}. The same can be said of the theoretical description of both H$_2$ \cite{yo2} and D$_2$ \cite{yo3}           
adsorbed on graphite. 

We expect then, that this intermolecular potential could describe reasonably 
the phases of D$_2$ on this novel surface. 
  
The
C-D$_2$ and H-D$_2$ substrate potentials were assumed to be of Lennard-Jones
type. Since the hybridization of the carbon atoms on graphane is $sp^3$
instead of the $sp^2$ one of graphene and graphite, one cannot use the same
parameters as in previous simulations  of adsorption on the latter 
substrates. We resorted
then to Ref. \cite{sp3}   were the C-C and H-H Lennard Jones
parameters for CH$_4$ (a compound where the carbon atoms have a 
$sp^3$
hybridization) were given. Then, the Lorentz-Berthelot combination rules were
applied, taking the corresponding $\epsilon$ and $\sigma$ D$_2$-D$_2$
values from Ref. \cite{uptake}.     The Lennard Jones parameters so
obtained are $\epsilon_{C-D_2}$ = 43.52 K, $\sigma_{C-D_2}$ = 3.2 \AA, 
$\epsilon_{H-D_2}$ = 13.42 K, and $\sigma_{H-D_2}$ = 2.83 \AA. This is our reference
set of interaction parameters, that from now on,
will be referred to as LJ1. Since we cannot be sure of the accuracy of the
approximation used (after all, graphane is not CH$_4$), we considered another set of
Lennard Jones parameters for the H-D$_2$ interaction (from now on referred to as LJ2). 
The basic idea is to check if the phase
diagram of D$_2$ on graphane is reasonable robust with respect to variations in the 
D$_2$-surface interaction. However, we only changed the H-D$_2$ parameters with respect to LJ1 because
the C atoms are not in direct contact with the D$_2$ molecules, and therefore their
influence on the adsorbed deuterium molecules should be smaller.
We derived this LJ2 potential from the same above mentioned parameters
for CH$_4$ (Ref. \cite{sp3}), but used the D$_2$-D$_2$ ones that result from applying
backwards the Lorentz-Berthelot rules to the C-H$_2$ interaction 
given in Ref. \cite{coleh2} for H$_2$ adsorbed on graphite. 
Obviously, the results derived for H$_2$ are valid for D$_2$, since 
the interaction potentials depend on the electronic structure of the 
atoms or molecules involved, and this is the same for both hydrogen isotopes. 
Using this last approximation, one gets
$\epsilon_{H-D_2} = $17.86 K and $\sigma_{H-D_2}$ = 2.56 \AA$ $ for this second 
interaction. Unfortunately, we cannot choose a 
potential set as been more accurate than the other, since there are not experimental  
data on the binding energy of D$_2$ on graphane to compare to. Our only goal is then to see if both phase diagrams are similar to each other. 
This would mean that  
we have a reasonable description of the experimental phases of deuterium on graphane, in the same  
way that we can describe accurately the behaviour of the same adsorbate on graphite using similar potentials \cite{yo2,yo3}.

The primary output of the application of the DMC
method  is the 
local energy, $E_L$, whose statistical mean for large enough
imaginary time 
corresponds  to the ground-state 
energy  of the system.~\cite{boro94} Explicitly,
\begin{equation}
E_L = \Psi({\bf r}_1, {\bf r}_2, \ldots, {\bf r}_N)^{-1} H \Psi({\bf r}_1, 
{\bf r}_2, \ldots, {\bf r}_N) \ ,  
\end{equation}
where 
\begin{equation}
H =   - \frac{\hbar^2}{2m}  \sum_{i=1}^N \bm{\nabla}_i^2 + \sum_{1=i<j}^{N}
V_{D_2-D_2} (r_{ij}) + \sum_{m,i=1}^{N_C,N}  V_{C-D_2} (r_{mi})  + 
\sum_{n,i=1}^{N_H,N}  V_{H-D_2} (r_{ni})
\label{hamilton}
\end{equation} 
is the Hamiltonian of the system.
$ \Psi $ stands for $\Psi_L$ or $\Psi_S$ 
depending on the phase considered. The local energy is our estimator for the ground state energy of a system described by a given {\em trial function}.
This is equivalent to say that we operate always at T = 0 K, temperature at which the free energy of a system 
equals its energy. If we compare then different arrangements of particles (described by different {\em trial functions}), 
the one whose energy per particle is 
minimum will be the ground state of the system as a whole. If we consider now arrangements with higher densities, we will eventually 
reach other stable phases, whose density limits will be determined via a standard double-tangent Maxwell construction
\cite{chandler}.


\section{RESULTS}

\begin{figure}
\includegraphics[width=1.0\linewidth]{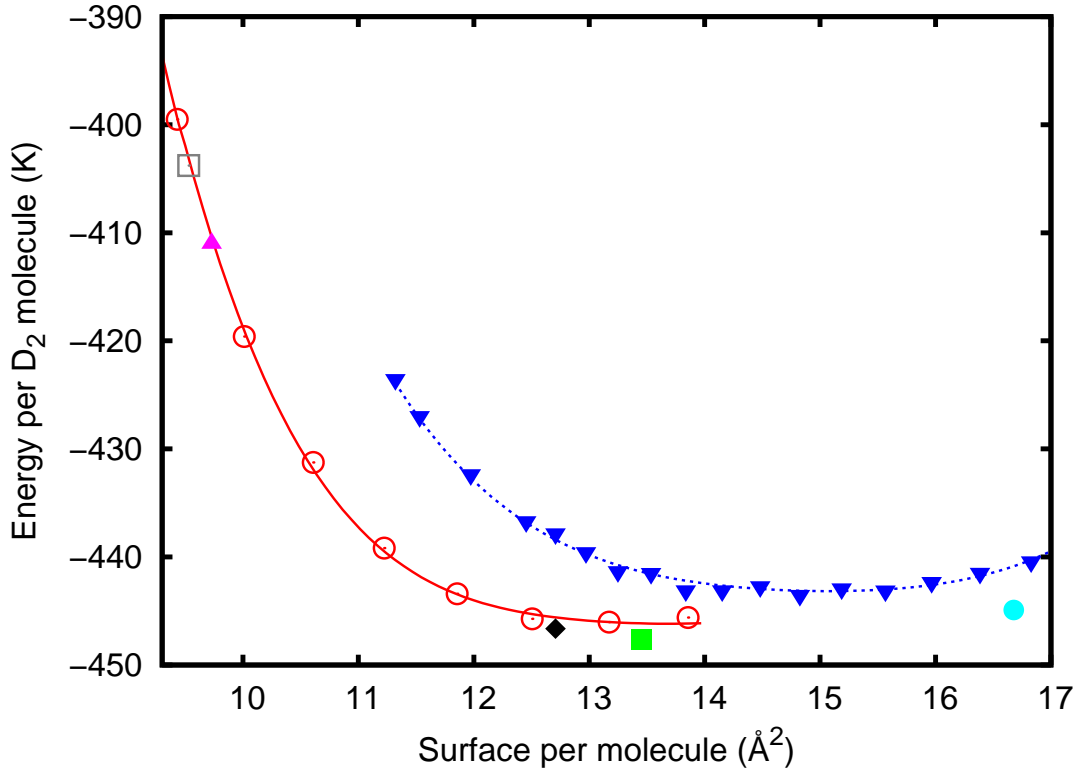}
\caption{\label{DLJ1}{Phase diagram for D$_2$ on graphane using the set of parameters LJ1. 
Full circles, $\sqrt3\times\sqrt3$. Full squares, $\delta$ phase. Solid
diamond, $\epsilon$ phase. Full triangle, 4/7 phase; open square, 7/12 commensurate solid. The liquid arrangements are 
represented by  inverted full triangles, while the open circles correspond to the 
incommensurate triangular solid.
The solid and dashed lines are fourth-order polynomial fits to their corresponding data sets. 
The error bars are of the same size of the symbols and are not displayed 
for simplicity.}}
\end{figure}

\begin{figure}
\includegraphics[width=1.0 \linewidth]{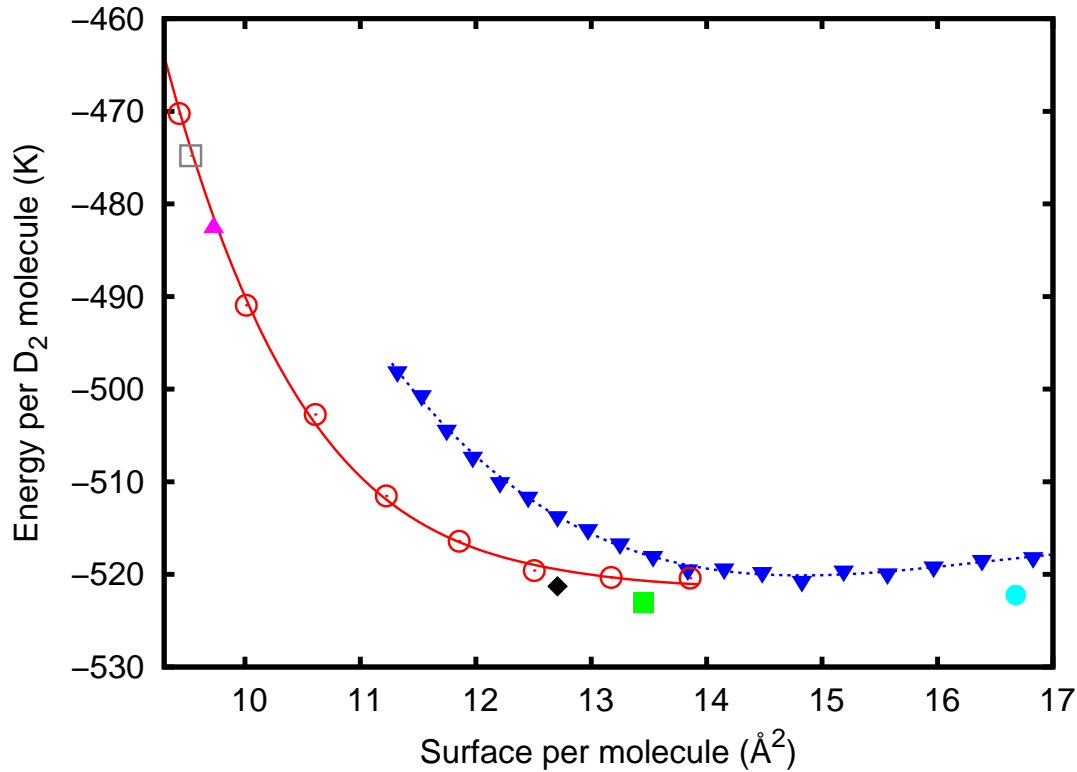}
\caption{\label{DLJ2}{Same as in figure
\ref{DLJ1}, but for the set of Lennard-Jones parameters LJ2}}
\end{figure}

The phase diagram of D$_2$ on graphane can be 
derived from the DMC energies reported in figure \ref{DLJ1}.
There, all the symbols correspond to simulation results both for a translational
invariant system (liquid, inverted triangles) and to different 
two-dimensional solids.
We plotted the energy per D$_2$ molecule versus the surface area, which is  
the inverse of the deuterium surface density. In that way, to perform the necessary
double-tangent Maxwell constructions to determine the stability regions of the different phases 
is straightforward.   
The solid arrangements considered were the standard triangular incommensurate phase, and the same 
commensurate structures taken into account in a previous calculation of D$_2$ on graphene 
($\sqrt 3 \times \sqrt 3$, $\delta$, and $\epsilon$ phases).~\cite{yo3}  
Those registered phases were taken as such with respect to the 
projections of the carbon atoms on the $z$ = 0 plane, projections that
form a honeycomb lattice. 
We tried also some structures that were commensurate 
with respect to the atomic hydrogen triangular lattice, taking as a model the ones proposed for 
a second layer of $^4$He on graphene,~\cite{yo4} i.e., the $4/7$ and $7/12$ phases. 
That system could be considered analogous to the one in the present work because a second $^4$He layer rests also on top of a 
triangular helium substrate. 
Our present results show that both the 4/7 and 7/12 structures
have similar energies per hydrogen molecule than 
their incommensurate counterparts at the same densities (see figures \ref{DLJ1} and \ref{DLJ2}), so there
is no way to know if they are separate phases.
It is worth noticing that the graphane unit cell that
builds up the entire structure is bigger than that of graphene. This means that
the corresponding adsorbate densities are lower than for a similar arrangement in graphene. 
For instance, a structure equivalent to the   
 $\sqrt 3 \times \sqrt 3$ solid in C-graphane has a density of  
0.0600 \AA$^{-2}$ instead of the value 0.0636 \AA$^{-2}$ found in graphene and graphite.

\begin{figure}
\includegraphics[angle=-90,width=1.0 \linewidth]{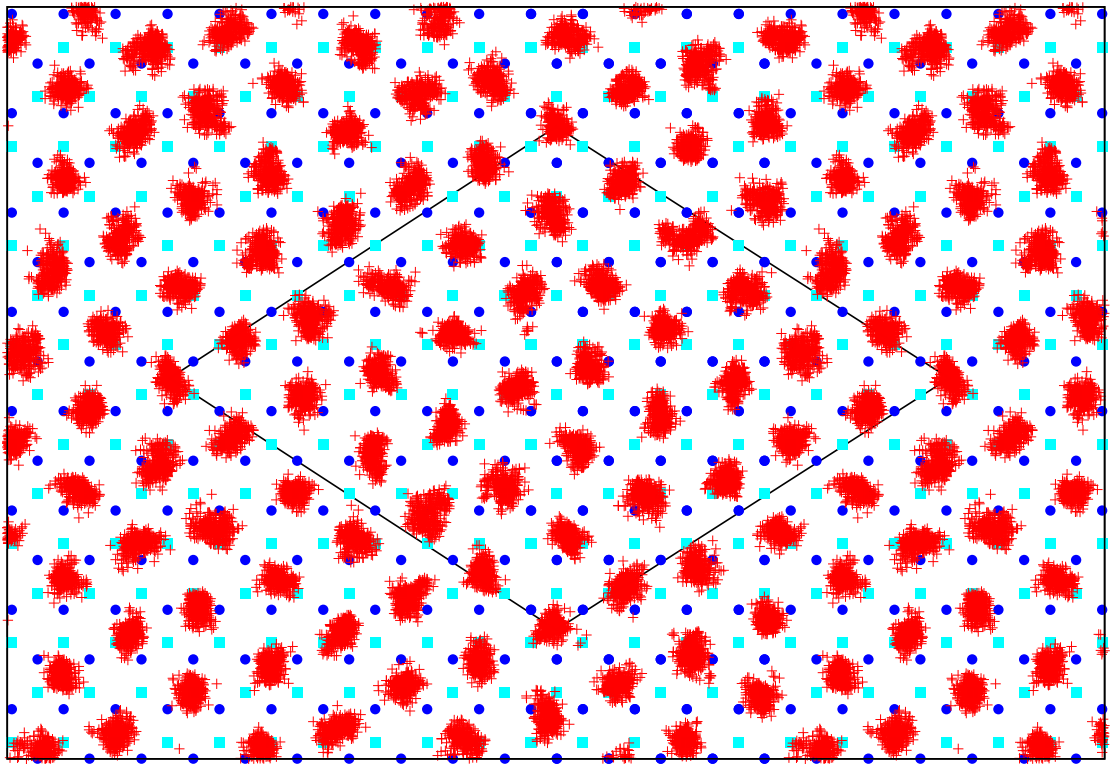}
\caption{\label{manchas_delta}{Sketch of the $\delta$ structure. Solid smudges are the result of displaying 300 sets of deuterium coordinates represented as crosses. 
Solid circles are the projection on the $z=0$ plane of the positions of the carbon atoms bound to the upper H atoms in the C-graphane structure. Solid squares
represent the carbon atoms bound to the bottom hydrogens in the skeleton. The big diamond is the unit cell for this arrangement. 
}}
\end{figure}

In figure \ref{DLJ1}, all the calculations were performed using the LJ1 set of Lennard-Jones
parameters.  
To check the influence of the adsorbate-surface interaction in the 
phase diagram, we 
used the alternative LJ2 potential. Those results are displayed in figure \ref{DLJ2}. 
The obvious conclusion from figure \ref{DLJ1} and figure \ref{DLJ2} is that, irrespectively of
the Lennard-Jones parameters  employed, and in the density range represented in both figures, 
the structure with lowest energy per particle for D$_2$ 
on C-graphane is a $\delta$ commensurate solid,
lower than the corresponding to a $\sqrt 3 \times \sqrt 3$ commensurate structure, and lower than for a liquid arrangement.
The corresponding energies for each phase are listed in 
table \ref{binding}. $E_0$ stands for the minimum energy per particle in the liquid phase, obtained 
from a  
fourth-order polynomial fit to the energies per particle displayed in figure \ref{DLJ1} and figure \ref{DLJ2}.
The binding energy of a single 
D$_2$ molecule on top of C-graphane surface is also given. This allows us to say that all the two dimensional 
adsorbed phases are less stable than their counterparts on graphene. 
The  $\delta$ structure is sketched in figure \ref{manchas_delta}. The big diamond displayed is its
unit cell, comprising 31 molecules. Four of these cells can be accommodated in a rectangular simulation cell 
of dimensions 38.0055 $\times$ 43.8890 \AA$^{2}$. This cell is big enough to prevent any size effects to appear.  
 We did not display the $\sqrt 3 \times \sqrt 3$ solid  since
it is a standard well known arrangement (see for instance the same structure on graphite in Ref. \cite{colebook}).    
The same can be say of the incommensurate triangular solid (see below).

\begin{figure}
\includegraphics[angle=-90,width=1.0 \linewidth]{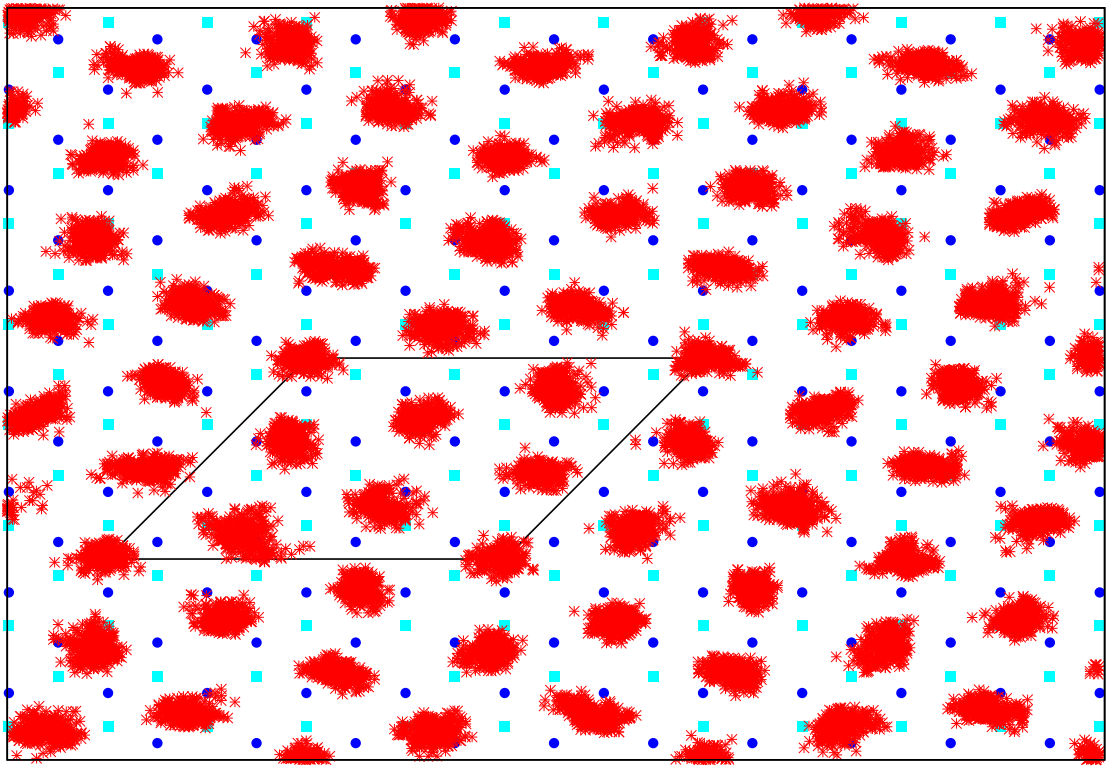}
\caption{\label{manchas_epsilon}{Same as in figure \ref{manchas_delta} for the $\epsilon$ arrangement. The rhomboid represents the unit cell.}}
\end{figure}

\begin{table}
\caption{\label{binding}{Energies in the infinite dilution limit, $E_{\infty d}$. Energies per molecule at the minima of the liquid curves in figure \ref{DLJ1} and figure \ref{DLJ2}, $E_0$ . 
Third, fourth and fifth columns show the adsorption energies of the liquid, $\sqrt 3\times \sqrt 3$ and $\delta$ phases respectively with respect the infinite dilution limit. For comparison, the same results for graphene 
\cite{yo3} are also included. }}

\begin{tabular}{cccccc}
&$E_{\infty d}(K)$&$E_0(K)$&$\left( E_0-E_{\infty d}\right)(K)$&$\left(E_{\sqrt3\times\sqrt3}-E_{\infty d}\right)(K)$&
$\left(E_\delta-E_{\infty d}\right)(K)$ \\
\hline
LJ1-graphane&-407.6330$\pm$0.0001&-443.3$\pm$0.3&-35.6$\pm$0.3&-37.279$\pm$0.006&-40.01$\pm$0.02\\
LJ2-graphane&-484.0974$\pm$0.0001&-520.1$\pm$0.3&-36.0$\pm$0.3&-38.121$\pm$0.006&-38.90$\pm$0.02\\
graphene&-464.87$\pm$0.06&-497.2$\pm$0.9&-32.3$\pm$0.9&-43.66$\pm$0.06&-40.75$\pm$0.07\\

\end{tabular}

\end{table}


On increasing the D$_2$ density, the next stable phase will be  the $\epsilon$
registered phase of density 0.0787 \AA$^{-2}$, and represented by a  
solid diamond both in figure \ref{DLJ1} and
figure \ref{DLJ2}). Its sketch is given in figure \ref{manchas_epsilon}, that displays its unit cell containing
seven molecules.  We can accommodate 112 D$_2$ molecules of this arrangement in a rectangular 
simulation cell of 40.5392 $\times$ 35.1112 \AA$^{2}$, also big enough to avoid any kind of size effects. 
A piece of that simulation cell, enough to show the primitive unit, is displayed in figure \ref{manchas_epsilon}.  
Since the $\delta$ and $\epsilon$ structures are represented by a single density, 
the double tangent Maxwell construction between them is simply the line that joints 
both symbols. In both figures and in table \ref{d2}, we can see that the
$\epsilon$ solid is more stable that an incommensurate arrangement of the same
density. 
This means that
upon a density increase, the  phase diagram for D$_2$ on C-graphane would
proceed through the sequence $\delta$ $\rightarrow$ $\epsilon$
$\rightarrow$ incommensurate triangular solid. The lowest density
of the incommensurate lattice 
(obtained from a Maxwell construction
between the $\epsilon$ and this structure) was 0.084 $\pm$ 0.002 \AA$^{-2}$
for both series of Lennard-Jones parameters.

\begin{table*}
\caption{\label{d2}{Energies per molecule and densities of the different phases 
of D$_2$ on graphane.}}
\begin{indented}
\item[]
\begin{tabular}{cc|cc|cc}
&&LJ1&&LJ2&\\
Phases&Density ($\AA^{-2}$) &Energy(K)&Energy(K)&Energy(K)&Energy(K)\\
\hline
Liquid&&-443.3$\pm$0.3$^{\rm a}$&&-520.1$\pm$0.3$^{\rm b}$\\
$\sqrt3\times\sqrt3$&0.0600&-444.912 $\pm$0.006&-440.8$\pm$0.3$^{\rm c}$&-522.218 $\pm$ 0.006&-518.3$\pm$0.3$^{\rm c}$\\
$\delta$&$0.0743$&-447.64$\pm$ 0.02&-446.0$\pm$0.1$^{\rm d}$&-523.00 $\pm$0.02&-520.5$\pm$0.3$^{\rm d}$\\
$\epsilon$&$0.0787$&-446.64$\pm$ 0.02&-445.7$\pm$0.1$^{\rm d}$&-521.29$\pm$ 0.02&-519.6$\pm$0.2$^{\rm d}$\\
\end{tabular}

\item[]$^{\rm a}${At density 0.067$\pm$0.001 $\AA^{-2}$}
\item[]$^{\rm b}${At density of 0.055$\pm$0.001 $\AA^{-2}$}
\item[]$^{\rm c}${comparison with the liquid phase.}
\item[]$^{\rm d}${comparison with the incommensurate solid.}

\end{indented}
\end{table*}

\section{CONCLUSIONS}

We calculated the phase diagram of D$_2$ on
C-graphane, a novel substance that has been experimentally
realized.  
Both the structure
of the compound and all the interactions between the different parts of the
system were taken to be as much realistic as possible. This means that the
results of our work could be checked against experimental data in the
future. The fact that both the stable phases and their density limits were
unchanged by modifications of the surface-deuterium interaction potentials 
makes us confident in the reliability of the method and in our conclusions.          
Since we have no experimental data to compare to, we cannot reach any conclusion
about the deuterium adsorption energies.  In this, we are at disadvantage
with the case of graphene, for which we do not have experimental data
either, but whose energies could be compared to those of graphite, a close
related compound.    

Our results also indicate that the ground state of
deuterium adsorbed on graphane is the registered phase $\delta$, 
what makes D$_2$ on graphane different from H$_2$ on graphane \cite{JLTP_ccc}, or from D$_2$ or any other quantum gas 
on graphene \cite{yo,yo2,yo3}, 
where the ground states were $\sqrt 3 \times \sqrt 3$ arrangements.
This is also at odds with some
recent results for
$^4$He on graphane.~\cite{reatto} Those indicate that  the ground
state of $^4$He  on graphane was a liquid, and that a registered
phase analogous to the $4/7$ structure  was also stable. We did not found
that the energy per molecule of that phase were appreciably different than the corresponding to an incommensurate triangular
phase of the same density for D$_2$. 
In any case,
the differences between the phase diagrams on graphene
and graphane could make the last one an interesting object of experimental
study in the  future.    
    
\ack
We acknowledge partial financial support from the Junta de
de Andaluc\'{\i}a Group PAI-205, Grant No. FQM-5987, MICINN 
(Spain) Grants No. FIS2010-18356 and FIS2011-25275, and Generalitat de Catalunya
Grant 2009SGR-1003.

\end{document}